\documentclass[apj]{emulateapj}

\newcommand{\msunh}{\>h^{-1}\rm M_\odot}

\newcommand{\mpch}{\>h^{-1}{\rm {Mpc}}}

\newcommand{\kmsmpc}{\>{\rm km}\,{\rm s}^{-1}\,{\rm Mpc}^{-1}}

\usepackage{xcolor}

\shorttitle{Revealing the cosmic web dependent halo bias} \shortauthors{Yang et al.}

\begin{document}
            

\title{Revealing the cosmic web dependent halo bias}
    
\author{Xiaohu Yang\altaffilmark{1,2,3}, Youcai Zhang\altaffilmark{4},
  Tianhuan Lu\altaffilmark{5}, Huiyuan Wang\altaffilmark{6}, Feng
  Shi\altaffilmark{4}, Dylan Tweed\altaffilmark{1}, Shijie
  Li\altaffilmark{1}, Wentao Luo\altaffilmark{1}, Yi
  Lu\altaffilmark{4}, Lei Yang\altaffilmark{1}}

\altaffiltext{1}{Department of Astronomy, Shanghai Jiao Tong
  University, Shanghai 200240, China; E-mail: xyang@sjtu.edu.cn}

\altaffiltext{2}{IFSA Collaborative Innovation Center, Shanghai Jiao
  Tong University, Shanghai 200240, China}

\altaffiltext{3}{Tsung-Dao Lee Institute, Shanghai 200240, China}

\altaffiltext{4}{Shanghai Astronomical Observatory, Nandan Road 80,
  Shanghai 200030, China}

\altaffiltext{5}{Zhiyuan College, Shanghai Jiao Tong University,
  Shanghai 200240, China}

\altaffiltext{6}{Key Laboratory for Research in Galaxies and
  Cosmology, Department of Astronomy, University of Science and
  Technology of China, Hefei, Anhui 230026, China}


\begin{abstract}
  Halo bias is the one of the key ingredients of the halo models. It
  was shown at a given redshift to be only dependent, to the first
  order, on the halo mass. In this study, four types of cosmic web
  environments: clusters, filaments, sheets and voids are defined
  within a state of the art high resolution $N$-body simulation.
  Within those environments, we use both halo-dark matter
  cross-correlation and halo-halo auto correlation functions to probe
  the clustering properties of halos.  The nature of the halo bias
  differs strongly among the four different cosmic web environments we
  describe.  With respect to the overall population, halos in clusters
  have significantly lower biases in the
  {$10^{11.0}\sim 10^{13.5}\msunh$} mass range. In other environments
  however, halos show extremely enhanced biases up to a factor 10 in
  voids for halos of mass {$\sim 10^{12.0}\msunh$}. Such a strong
  cosmic web environment dependence in the halo bias may play an
  important role in future cosmological and galaxy formation
  studies. Within this cosmic web framework, the age dependency of
  halo bias is found to be only significant in clusters and filaments
  for relatively small halos $\la 10^{12.5}\msunh$.
\end{abstract}




\section{Introduction}

In the current paradigm of structure formation, the virialized dark
matter halos are considered to be the building blocks of the mass
distribution in the universe. The structure and number distribution of
dark matter halos, as well as their formation histories and clustering
(bias) properties, are the main ingredients of halo models.  Among
these properties, theoretical halo bias models have been derived
either analytically using the (extended) Press-Schechter formalism
\citep[e.g.][]{Press1974, Bardeen1986, Mo1996, Sheth1999, Sheth2001},
or formulated empirically from numerical simulations
\citep[e.g.][]{Jing1998, Seljak2004, Tinker2005, Pillepich2010,
  Tinker2010}.  So far, at a given redshift, the halo bias appears to
have a first order dependance only on halo mass with more massive
halos tending to be more strongly clustered.  This mass dependence has
played crucial roles both in cosmological probes using the clustering
measurements of clusters or groups \citep[e.g.][]{Bahcall2003,
  Yang2005}, and in galaxy formation studies such as understanding the
correlation functions of dark matter and galaxies via so called halo
models \citep[e.g.][]{Cooray2002}, halo occupation models
\citep[e.g.][]{Jing1998b, Berlind2002}, and conditional luminosity
functions \citep[e.g.][]{Yang2003, Bosch2003, Yang2012}.

Apart from the mass dependence, many studies in recent years have
tried to reveal additional dependences, among which the most notable
is the {\it halo assembly bias} \citep[e.g.][] {Sheth2004, Gao2005,
  Gao2007}. The assembly bias was first observed among halos of
similar masses in simulations where halos that formed earlier are more
strongly clustered than those that formed later \citep{Gao2005}.
Different studies shed light on additional second or third order
dependencies on spin, shape, substructures and halo trajectory
\citep[e.g.][] {Wechsler2006, Gao2007, Li2008, Dalal2008, Wang2011,
  More2016, Mao2017}.  In observations, \citet{Yang2006} was the first
claiming detection of an age dependence in the halo bias from
clustering measurements of galaxy groups in the 2dFGRS
\citep{Colless2001}. Additional confirmed findings were made from
various observations \citep[e.g.][]{Blanton2007, Swanson2008,
  Wang2008, Wang2013, Deason2013, Lacerna2014, Miyatake2016}.
However, null detection was also reported \citep[e.g.][]{Lin2016,
  Zu2017}. By means of galaxy bias models or halo occupation
distribution and conditional luminosity function models, the
clustering properties of galaxies have been extensively used in
cosmological probes and galaxy formation constraints. The impact of
the assembly bias is thus an important issue that one needs to take
into account. For this reason the degree of assembly bias that is
transferred to galaxies and its impact on cosmology and galaxy
formation have been extensively discussed \citep[e.g.][] {Croton2007,
  Reed2007, Zu2008, Jung2014, Zentner2014, Hearin2015, Tonnesen2015,
  Hearin2016}.  So far, these studies have shown that the impact of
assembly bias on galaxy clustering properties is quite trivial.

In addition to the halo structures, numerical simulations and large
galaxy redshift surveys have also shown striking structures: clusters,
filaments, sheets, and voids. These diversity of structures are best
referred as the cosmic web. From a dynamical point of view, dark
matter flows out of the voids, accretes onto the sheets, collapses
into filaments and finally accumulates in clusters. In this picture,
the assembly histories of halos as well as the galaxies formed in them
are expected to be affected by the large scale environment. There are
different approaches in literature to quantify the cosmic web
environments, among which the most straightforward one is using
Hessian matrix.  \citet{Hahn2007a, Hahn2007b} have quantified the
cosmic web environments using the Hessian matrix of the potential
field where, according to the number of positive eigenvalues, a region
was classified as belonging either to a cluster, a filament, a sheet, or
a void environment. The only free parameter in this analysis is the
smoothing length of the density field. Similar probes were carried out
by \citet{Aragon2007a, Aragon2007b}, who computed the Hessian matrix
based on the density field constructed with a Delaunay triangulation
field estimator \citep[see also][ and references therein]{Zhang2009}.

Over the past decade, numerical simulations have revealed that the
properties of halos have dependences on the large scale environments
in which they reside \citep{Sheth2004, Hahn2007b, Jung2014,
  Fisher2016, Borzy2017, Lee2017, Paranjape2017}.  \citet{Sheth2004}
showed that halos in dense environment form slightly earlier than
halos of the same mass in less dense environment. \citet{Hahn2007b}
claimed that low mass halos in the four different environments have
significantly different assembly histories. Particularly, low mass
halos at fixed mass tend to be older in clusters and younger in voids.
Using the Millennium simulation, \citet{Fisher2016} show that at fixed
halo mass, the clustering properties of halos dependent on the cosmic
web environments significantly.  \citet{Borzy2017} investigated the
origin of halo assembly bias, using 7 zoom-in simulations of
{$\mathcal{O}\left( 10^{11}\msunh\right)$} halos. They concluded that
halo assembly bias originates from quenching halo growth by tidal
interaction during the formation of nonlinear structures in the cosmic
web.

Several observational studies also endeavored to investigate the
dependence of either galaxy or halo properties within different large
scale environments. Using galaxy groups that are associated with dark
matter halos, \citet{Wang2009} proposed a sophisticated and robust way
to reconstruct the density field of Universe. Based on the matter
density field constructed from galaxy groups, \citet{Zhang2009}
classified the groups/galaxies in the SDSS observation into different
cosmic web environments and probed the alignment signals of
galaxies. In addition to the cosmic web classification,
\citet{Wang2012} used the galaxy groups in the SDSS observation to
reconstruct the mass density, tidal and velocity fields in the local
Universe. Thus obtained mass density field was used to perform the
constrained simulation of the local universe in \citet{Wang2016},
where they found that the red fraction of galaxies in four different
cosmic web environments do show very different behaviors.  Apart from
the SDSS observation, using the GAMA survey \citep{Driver2011},
\citet{Eard2015} computed the tidal tensor from a smoothed galaxy
density field and classified galaxies into different cosmic web
environments. They showed that there is no significant influence of
the cosmic web on galaxy luminosity functions. Using the same sets of
data, a recent study by \citet{Tojeiro2016} indicates that low mass
halos in clusters are older than halos of similar mass in
voids. However, the trend is reversed for high mass halos, with halos
in clusters being younger than halos in voids. Their work provides the
first direct observational evidence for halo assembly bias in
connection with strong tidal interactions.  Note however, since
galaxies are biased tracers of dark matter, caution should be
exercised when interpreting these observational results where the
cosmic web environments are classified according to the smoothed
galaxy density field.

In this study, with the help of a large N-body simulation, we set out
to measure the halo clustering properties in different cosmic web
environments. Here we mainly focus on the large scale behaviors of the
clustering measures, i.e., the biases of the halos.  The purpose of
this work are two folded: (1) to see if the halo biases have
significant cosmic web environmental dependence, which might be useful
for cosmological probes \citep[e.g.][]{Hamaus2014, Dai2015,
  Hamaus2016}, and (2) to see if the age dependent halo assembly bias
can be explained in terms of the cosmic web environmental dependence
\citep[e.g.][]{Borzy2017}.

This paper is organized as follows.  Section \ref{sec_data} gives a
detailed description of the data and method we used in this study.  In
Section~\ref{sec_clustering}, we investigate the clustering properties
of dark matter halos in different comic web environments.  In
Section~\ref{sec_age}, we explore the age dependence of the halo
biases in those environments. We summarize our results in
Section~\ref{sec_conclusion}.  Throughout the paper we adopt a
$\Lambda$CDM cosmology with parameters that are consistent with the
fifth-year data release of the WMAP mission \citep[][hereafter WMAP5
cosmology]{Dunkley2009}; $\Omega_{\rm m} = 0.258$,
$\Omega_{\Lambda} = 0.742$,$\Omega_{\rm b} = 0.044$,
$h=H_0/(100 \kmsmpc) = 0.719$, $n=0.963$ and $\sigma_8 = 0.796$.

\begin{figure*}
\center
\vspace{0.5cm}
\includegraphics[height=10.0cm,width=10.5cm,angle=0]{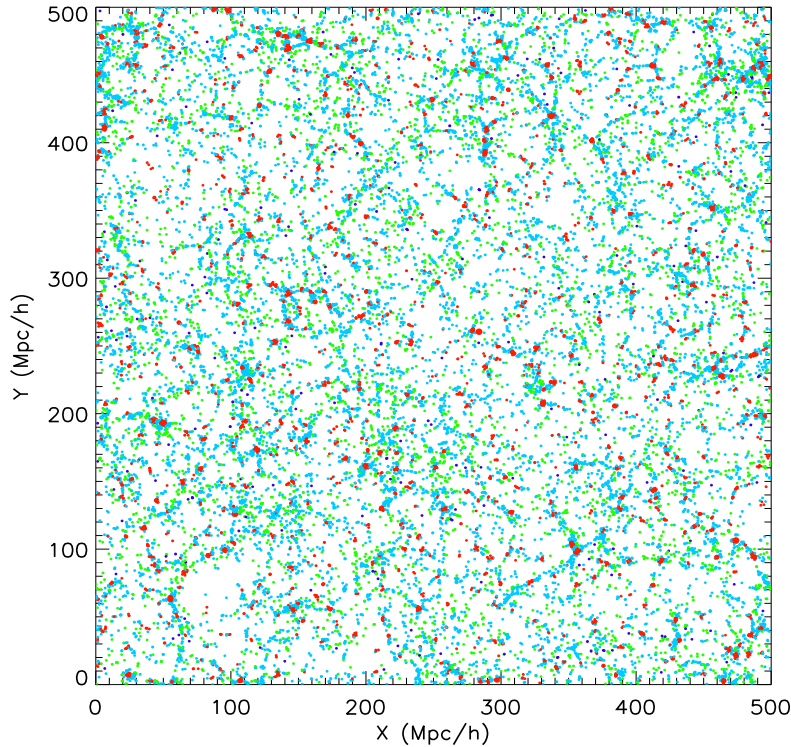}
\caption{Projected distribution of halos within a $2\mpch$ thick
  slice. For clarity, only randomly selected 20\% halos with masses
  larger than $10^{10} \msunh$ are shown and color coded in red, cyan,
  green and blue colors accordingly to their cosmic web environments:
  clusters, filaments, sheets and voids, respectively. }
\label{fig:slice}
\end{figure*}

\section{Data and method}
\label{sec_data}

\subsection{Halos in the ELUCID simulation}
\label{sec:simulation}

In this study we use dark matter halos extracted from the ELUCID
(Exploring the Local Universe with re-Constructed Initial Density
field) simulation.  This simulation which evolves the distribution of
$3072^{3}$ dark matter particles in a periodic box of $500 \mpch$ on a
side was carried out in the Center for High Performance Computing,
Shanghai Jiao Tong University. The simulation was run with {\tt
  L-GADGET}, a memory optimized version of {\tt GADGET2}
\citep{Springel2005b}. The cosmological parameters adopted by this
simulation are consistent with WMAP5 results with a mass per particle
of $3.0875\times10^{8}\msunh$.

Even though this is not relevant to this particular study, the ELUCID
simulation is a reconstruction of the mass density field extracted
from the galaxy \citep[e.g.][]{Blanton2005} and group
\citep[e.g.][]{Yang2007} distribution in the north galactic pole
region of the SDSS Data Release 7 \citep{Abazajian2009}. This density
field has been used to constrain the initial conditions using a
Hamiltonian Markov Chain Monte Carlo method with particle mesh
dynamics \citep[ELUCID I:][]{Wang2014}. The genesis of the ELUCID
simulation in particular and its basic properties such as the output
power spectrum and halo mass functions are described in detail in
\citet[][ELUCID III]{Wang2016}.

Dark matter halos have been identified in the ELUCID simulation with
the friends-of-friends algorithm. We have used a linking length
of $0.2$ times the mean particle separation. Only halos containing at
least $20$ particles are used for our study.  The dark matter halo
mass function of this simulation at redshift $z=0$ is in very good
agreement with the theoretical predictions of \citet{Sheth2001} and
\citet{Tinker2008}.

\subsection{Separating halos into different cosmic web types}
\label{sec:cosmicweb}

\begin{figure}
\center
\vspace{0.5cm}
\includegraphics[height=7.3cm,width=7.5cm,angle=0]{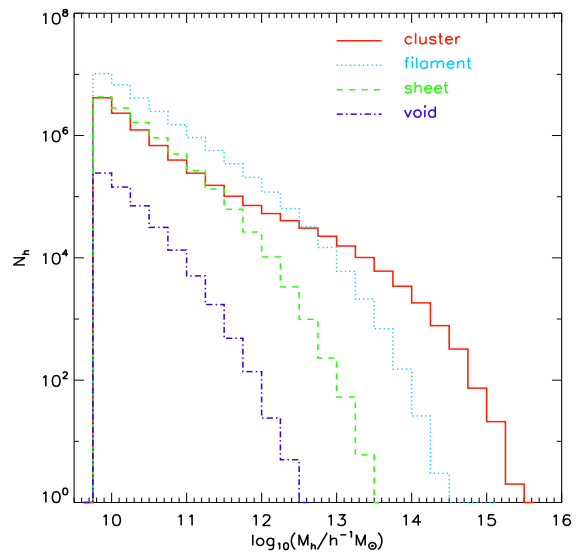}
\caption{Mass distribution of halos.  Red, cyan, green and blue
  lines indicate the mass distributions for halos in cluster,
  filament, sheet and void environments, respectively.}
\label{fig:mass}
\end{figure}
\begin{figure*}
\center
\vspace{0.5cm}
\includegraphics[height=10.0cm,width=10.0cm,angle=0]{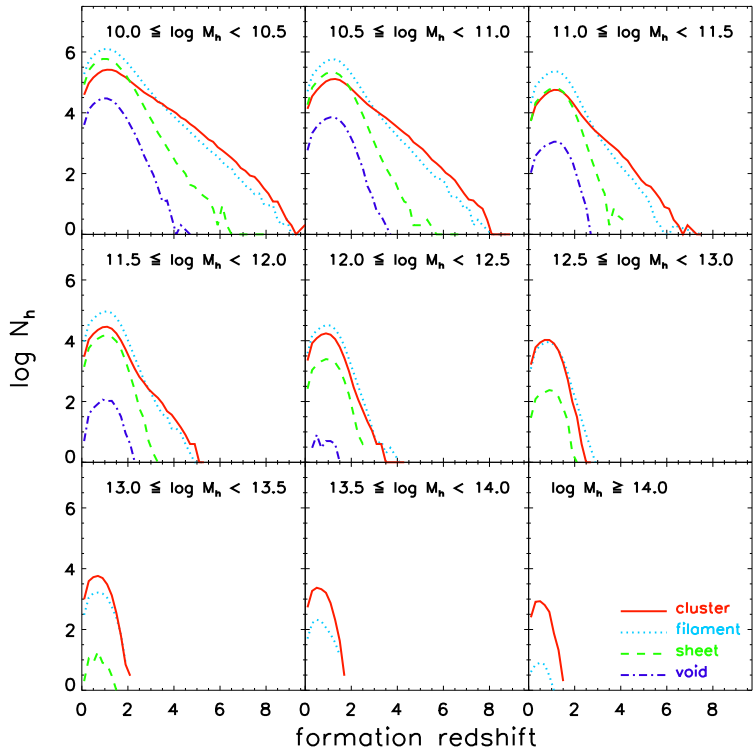}
\caption{Formation redshift of halos within different cosmic web
  environments. Each panel correspond to a specific halo mass bin.
  Red, cyan, green and blue lines indicate the formation redshift
  distributions for halos in cluster, filament, sheet and void
  environments, respectively. }
\label{fig:age}
\end{figure*}

In order to probe the clustering properties of halos within different
cosmic web environments, we first classify the dark
matter halos' environment using the method by \citet{Zhang2009}. This method is
based on the Hessian matrix of the smoothed density field. This matrix
is defined as
\begin{equation}
  H_{i,j} = \frac {\partial^2 \rho_s} {\partial x_i \partial x_j},
\end{equation}
where the Hessian matrix indices $i$ and $j$ and the corresponding
coordinates $x_i$ and $x_j$ can take values from the cartesian
coordinates along the box axis $x$, $y$ and $z$. The smoothed density
field $\rho_s$ is calculated using a Gaussian filter with a smoothing
scale $R_s=2.1 \mpch$ \citep{ Hahn2007a, Zhang2009}. The 3 eigenvalues
$\lambda_{[1,2,3]}$ of the Hessian matrix $H_{i,j}$ are then
calculated at the position of each halo. We adopt the convention where
$\lambda_1\leq\lambda_2\leq\lambda_3$. The number of negative
eigenvalues of $H_{i,j}$ can be used to classify the environment in
which the halo resides as follows
\citep[see][]{Zhang2009}:
\begin{description}
\item[cluster] a point where all three eigenvalues are negative,\\
  $\lambda_1\leq\lambda_2\leq\lambda_3<0$;
\item[filament] a point with 2 negative eigenvalues,\\
  $\lambda_1\leq\lambda_2<0<\lambda_3$;
\item[sheet] a point with only 1 negative eigenvalue,\\
  $\lambda_1<0<\lambda_2\leq \lambda_3 $;
\item[void] a point with no negative eigenvalues,\\
  $0<\lambda_1\leq\lambda_2\leq\lambda_3$.
\end{description}
Of the $48,129,323$ halos at redshift $z=0$ in the ELUCID simulation,
$9,529,790$ $(19.8\%)$ are located in clusters,
$27,439,946$ $(57.0\%)$ are located in filaments,
$10,649,415$ $(22.1\%)$ are located in sheets and
$510,172$ $(1.1\%)$ are located in voids.
Fig.~\ref{fig:slice} shows the projected distribution of dark matter
halos in the ELUCID simulation in a $500\mpch \times 500\mpch$ slice
of thickness $2\mpch$. Halos in clusters, filaments, sheets, and voids
are indicated with red, cyan, green and blue colors, respectively.

We show in Fig. \ref{fig:mass} the mass distributions of halos in
different cosmic web environments. Halos are the least numerous in the
voids and are more numerous in the filaments than in the sheets. 
In the clusters, the number of halos obeys a
different trend. At low mass ($M_{\rm h}<10^{11.5} \msunh$) end, this
number lies between those obtained for voids and sheets  and
at intermediate mass range ($10^{11.5}<M_{\rm h}<10^{12.5} \msunh$)
between those of sheet and filament numbers. As higher and higher
masses are explored, halos in clusters increasingly dominate the whole
population.

The formation of dark matter halos is a complex physical process which
can be characterized through a timescale: the halo formation time.
Among the various existing definition found in the literature, we use
the most common one which corresponds to the time at which the halo's
main branch assembled half of its present (redshift $z=0$) mass
$M_{\rm h}$.  Fig.~\ref{fig:age} shows the distribution of halos as a
function of their formation redshift. The different cosmic web
environments are distinguished and denoted by different colors. Each
panel corresponds to a different halo mass range as indicated. The age
distribution of halos are quite similar in the filaments, sheets and
voids. The halos in clusters however, show broader age distributions,
especially for low mass ones.  Nevertheless, the peak formation
redshifts of halos in all four cosmic web environments are similar,
which gradually change from $z\sim 1$ in low mass halos to $z\sim 0.5$
in massive halos.  In addition, we find that low mass halos
($M_{\rm h}<10^{11.0} \msunh$) in clusters (red) are slightly older
than halos of the same mass that reside in voids (blue).

\subsection{Cross correlation and auto correlation functions}
\label{sec:2PCF}

With all the halos being classified as part of different cosmic web
environments, we proceed to probe their clustering properties. The
first quantity we measure is the cross correlation function (CCF)
between halos and dark matter particles,
\begin{equation}
\xi_{\rm CCF} (r) = \frac{P_{\rm HD}(r)}{P_{\rm HR}(r)} -1\,,
\end{equation}
where $P_{\rm HD}(r)$ and $P_{\rm HR}(r)$ are the number of halo-dark
matter and halo-random pairs, respectively.  For our investigations,
the number of random points has been set to be the same as the number
of dark matter particles within the simulation box. Those points follow
a uniform distribution within the simulation volume.

In order to ensure self-consistency of the CCFs, we also measure the auto
correlation function (ACF) of dark matter halos, 
\begin{equation}
\xi_{\rm ACF} (r) = \frac{N_{\rm R}^2 P_{\rm HH}(r)}{N_{\rm H}^2P_{\rm
    RR}(r)} -1\,,
\end{equation}
where $P_{\rm HH}(r)$ and $P_{\rm RR}(r)$ are the number of halo-halo
and random-random pairs, $N_{\rm H}$ and $N_{\rm R}$ are the number of
halos and random points, respectively.

\section{The clustering properties of halos in different cosmic web
  environments}
\label{sec_clustering}

\begin{figure*}
\center
\vspace{0.5cm}
\includegraphics[height=10.0cm,width=10.0cm,angle=0] {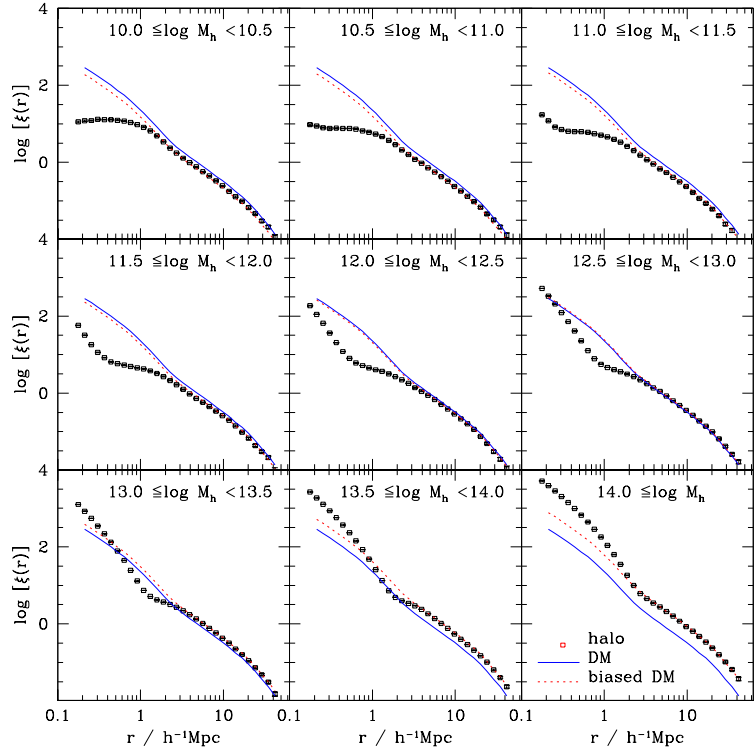}
\caption{Cross correlation functions between halos and dark matter
  particles. The different panels correspond to the same mass ranges
  as in Fig. \ref{fig:age}. Results are shown with open squares and
  error bars. For reference, the solid lines indicate the dark matter
  auto correlation in the ELUCID simulation and the dotted lines the
  same auto correlation multiplied by the bias factor of the halos in
  consideration. }
\label{fig:CCF1}
\end{figure*}
\begin{figure*}
\center
\vspace{0.5cm}
\includegraphics[height=10.0cm,width=10.0cm,angle=0]{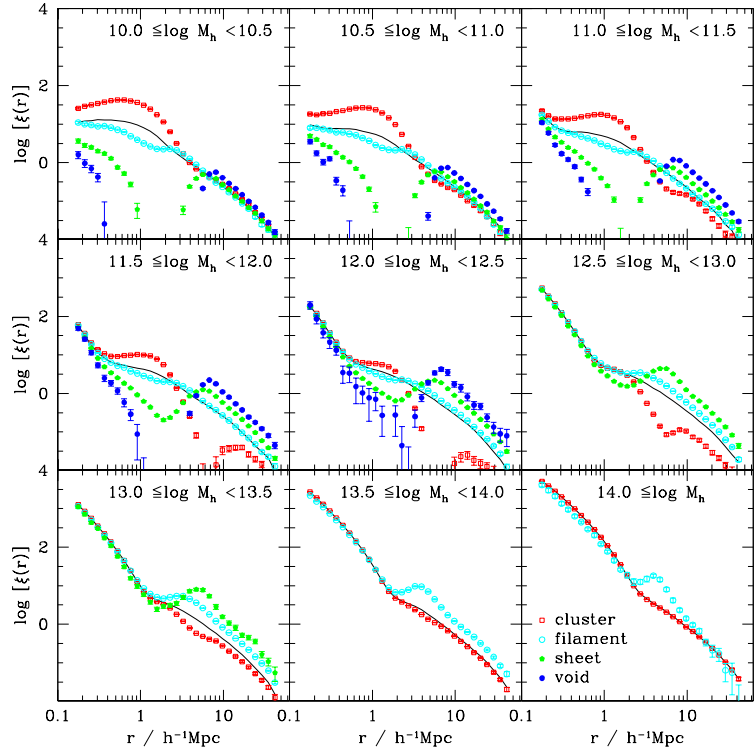}
\caption{Same as Fig. \ref{fig:CCF1}, but for halos in different
  cosmic web environments. In each panel, the open squares, open
  circles, stars and dots are results for halos in different cosmic web
  environments: clusters, filaments, sheets and voids, respectively. For
  comparison, results for the overall population are represented using
  a solid line. }
\label{fig:CCF2}
\end{figure*}

\subsection{Halo cross correlations}
\label{sec:ccf}

We first measure the CCFs between halos and dark matter particles in
the ELUCID simulation. We use the subsamples introduced in Fig
\ref{fig:age}, namely 9 mass ranges ($\Delta\log M_h =0.5$):
$10.0\le \log M_h <10.5$, $10.5\le \log M_h <11.0$, ...
$14.0\le \log M_h$.  The open squares shown in Fig. \ref{fig:CCF1} are
the CCFs measured from halos within these mass bins. The error bars
are obtained from 200 bootstrap re-samplings of the dark matter halos.
There are two features in these CCFs: 
\begin{enumerate}
\item more massive halos have overall stronger CCFs;
\item the 1-halo and 2-halo terms in the CCFs can be clearly separated
  by a change in the slope.
\end{enumerate}

For comparison, we also measured the auto correlation function (ACF)
of dark matter particles in the ELUCID simulation. It is shown in each
panel of Fig. \ref{fig:CCF1} as a solid line. Quite interestingly,
the CCF of halos with mass $\la 10^{13.5}\msunh$ is much smaller than
the ACF at the 1-halo and 2-halo transition scales.  This is caused by
the lack of 1-halo pairs in the CCF. On larger scales, $r\ga 3\mpch$,
the shape of CCF is quite consistent with the ACF.  Since massive
halos are more strongly clustered (biased) than low mass ones, to take
this bias into account in the ACF we use the theoretical predictions
obtained by \citet{Tinker2010}. The dotted line in each panel is the
ACF of dark matter particles multiplied by the median bias factor for
the halos in consideration. In general, we find that the biased ACFs
on large scales are in nice agreement with the CCFs.

Once we measured the CCFs of halos in different mass bins, we proceed
to measure the CCFs of subsamples of halos in different cosmic web
environments. We show in Fig \ref{fig:CCF2} the CCFs measured
separately from the four halo subsamples corresponding to their large
scale environments, namely: clusters (red open squares), filaments
(cyan open circles), sheets (green filled pentagons) and voids (blue
filled circles).  The CCFs of these subsamples show quite different
behaviors compared to the overall halo sample.  This is a clear
indication that a significant cosmic web environment dependence
exists, both on intermediate and large scales.  More precisely, this
figure reveals the following features:
\begin{enumerate}
\item On very small scales, where the CCFs are dominated by the 1-halo
  term, the CCFs of halos in different cosmic web
  environments follows the expected asymptotic trends. This behavior
  may not appear clearly for halos  $\la 10^{11.0}\msunh$ as the scale
  is limited to values $\ga 0.15\mpch$. Still it is quite distinct
  in more massive halos.
\item At intermediate scales between the 1-halo term region and
  $\sim 3\mpch$, some interesting variations are revealed especially
  for relatively low mass $\la 10^{12.5}\msunh$ halos. Halos in
  cluster, filament, sheet and void environments sequentially show
  suppressed CCFs. This feature is again quite expected, as the cosmic
  web environments themselves are defined according to the density
  field on such scales.
\item On large scales at $\ga 8\mpch$, however, an interesting and
  unexpected feature is found in the CCFs. The halos in cluster,
  filament, sheet and void environments sequentially show {\it
    enhanced} CCFs.  At any fixed mass, halos in the void region have
  the highest bias.
\end{enumerate}

\begin{figure*}
\center
\vspace{0.5cm}
\includegraphics[height=10.0cm,width=10.0cm,angle=0]{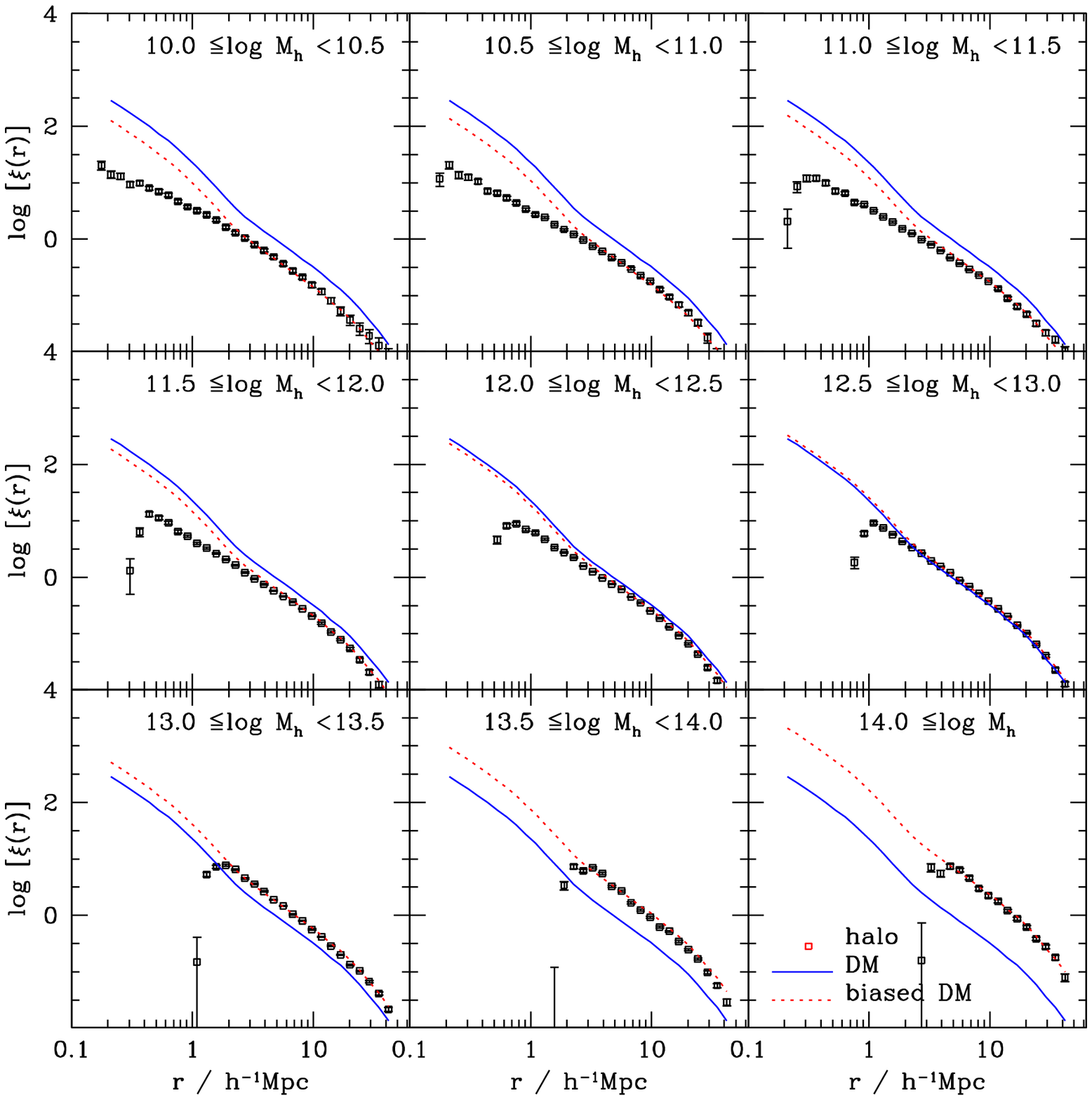}
\caption{Same as Fig, \ref{fig:CCF1}, but for the halo auto
  correlation functions. }
\label{fig:ACF1}
\end{figure*}
\begin{figure*}
\center
\vspace{0.5cm}
\includegraphics[height=10.0cm,width=10.0cm,angle=0]{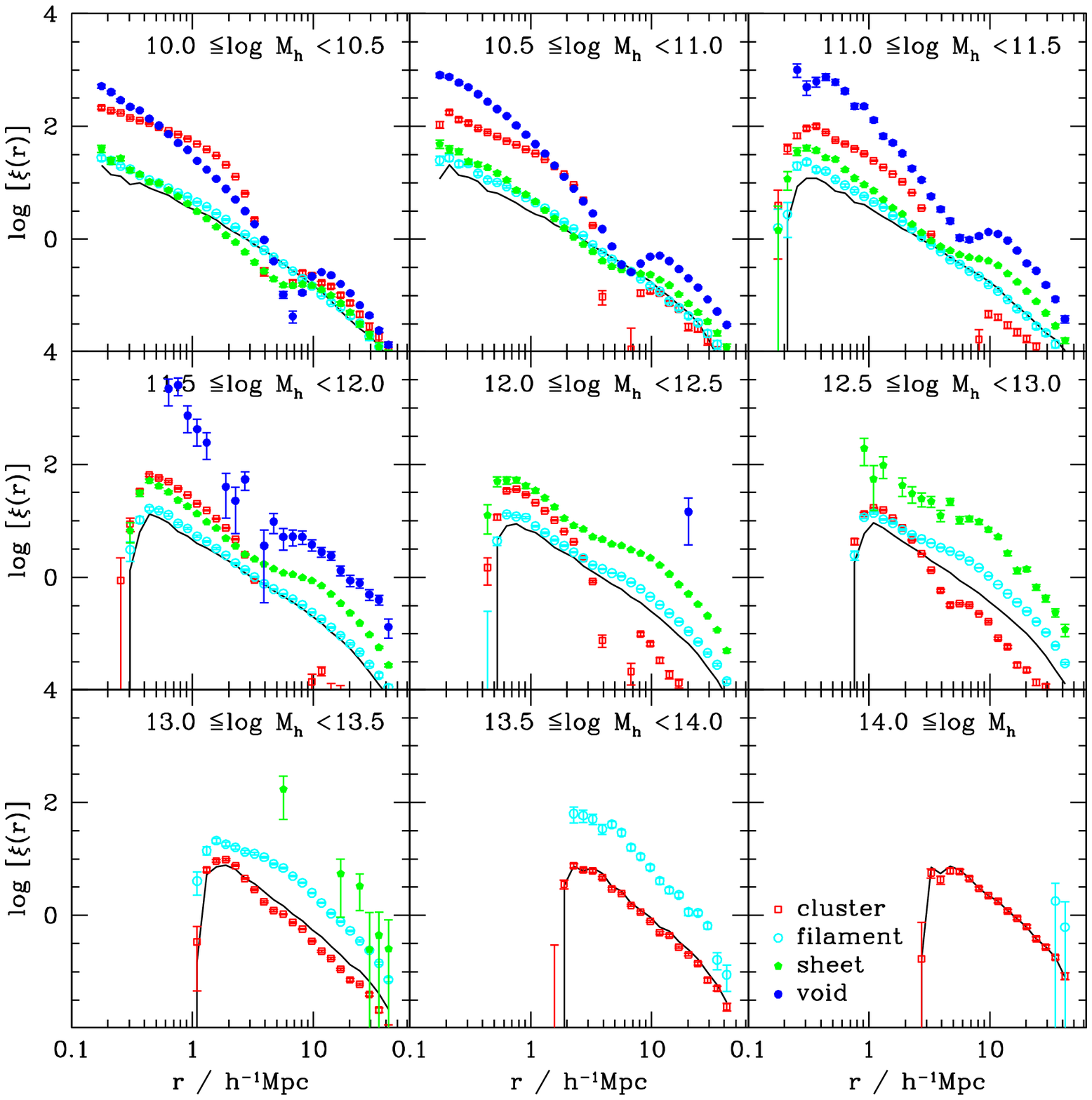}
\caption{Same as Fig, \ref{fig:CCF2}, but for the halo auto
  correlation functions. }
\label{fig:ACF2}
\end{figure*}

\begin{deluxetable*}{ccccccc}
  \tabletypesize{\scriptsize}
  \tablecaption{The biases of halos in different cosmic web environments.}
  \tablewidth{0pt}
  \tablehead{  (1) & (2) & (3) & (4) & (5) & (6) & (7) \\
  ID & $\log M_h$ & All & Cluster & Filament & Sheet & Void }

\startdata
  1 & 10.20  &   0.799 $\pm$  0.063  &   0.897 $\pm$  0.054  &   0.753 $\pm$  0.127  &   0.829 $\pm$  0.081  &   1.171 $\pm$  0.041 \\
  2 & 10.70  &   0.822 $\pm$  0.084  &   0.633 $\pm$  0.049  &   0.691 $\pm$  0.060  &   0.867 $\pm$  0.046  &   1.456 $\pm$  0.097 \\
  3 & 11.20  &   0.672 $\pm$  0.049  &   0.438 $\pm$  0.054  &   0.802 $\pm$  0.123  &   1.194 $\pm$  0.082  &   2.165 $\pm$  0.057 \\
  4 & 11.70  &   0.785 $\pm$  0.070  &   0.272 $\pm$  0.090  &   0.836 $\pm$  0.090  &   1.624 $\pm$  0.058  &   3.116 $\pm$  0.274 \\
  5 & 12.20  &   0.845 $\pm$  0.036  &   0.159 $\pm$  0.092  &   1.077 $\pm$  0.098  &   2.387 $\pm$  0.056  &   4.451 $\pm$  0.801 \\
  6 & 12.70  &   1.112 $\pm$  0.048  &   0.369 $\pm$  0.043  &   1.663 $\pm$  0.115  &   3.487 $\pm$  0.171  & -- \\
  7 & 13.20  &   1.248 $\pm$  0.075  &   0.896 $\pm$  0.066  &   2.459 $\pm$  0.156  &   4.692 $\pm$  0.703  & -- \\
  8 & 13.70  &   1.695 $\pm$  0.056  &   1.502 $\pm$  0.051  &   3.645 $\pm$  0.121  & --  & -- \\
  9 & 14.20  &   2.711 $\pm$  0.098  &   2.716 $\pm$  0.106  & --  & --  & -- \\
 \cline{1-7}\\
10 & 10.20  &   0.697 $\pm$  0.050  &   0.902 $\pm$  0.020  &   0.701 $\pm$  0.032  &   0.751 $\pm$  0.025  &   1.080 $\pm$  0.054 \\
 11 & 10.70  &   0.703 $\pm$  0.055  &   0.646 $\pm$  0.039  &   0.727 $\pm$  0.040  &   0.935 $\pm$  0.031  &   1.515 $\pm$  0.022 \\
 12 & 11.20  &   0.770 $\pm$  0.069  &   0.432 $\pm$  0.018  &   0.726 $\pm$  0.043  &   1.144 $\pm$  0.034  &   2.057 $\pm$  0.209 \\
 13 & 11.70  &   0.765 $\pm$  0.016  &   0.291 $\pm$  0.052  &   0.875 $\pm$  0.022  &   1.595 $\pm$  0.084  &   3.514 $\pm$  0.387 \\
 14 & 12.20  &   0.831 $\pm$  0.045  &   0.324 $\pm$  0.044  &   1.104 $\pm$  0.038  &   2.316 $\pm$  0.200  & -- \\
 15 & 12.70  &   1.035 $\pm$  0.039  &   0.522 $\pm$  0.049  &   1.642 $\pm$  0.076  &   3.416 $\pm$  0.367  & -- \\
 16 & 13.20  &   1.286 $\pm$  0.052  &   0.953 $\pm$  0.066  &   2.448 $\pm$  0.080  & --  & -- \\
 17 & 13.70  &   1.621 $\pm$  0.090  &   1.465 $\pm$  0.087  &   3.647 $\pm$  0.764  & --  & -- \\
 18 & 14.20  &   2.618 $\pm$  0.126  &   2.633 $\pm$  0.105  & --  & --   & -- 
\enddata

\tablecomments{Column (1): ID. Results listed in rows 1-9 are obtained
  from CCFs, 10-18 are obtained from ACFs.  Column (2): logarithm of
  the halo mass. Column (3): average biases of halos of overall
  population. Columns (4)-(7): average biases of halos in the cluster,
  filament, sheet and void environments, respectively. } \label{tab:bias}
\end{deluxetable*}

\subsection{Halo auto correlations}
\label{sec:acf}

The clustering strengths of large scale CCFs in different cosmic web
environments obviously contradict our naive expectation that halos in
voids should be less clustered. To test the robustness of
these findings, we proceed to calculate the halo ACFs.

Similar to the CCFs, we first measure the ACFs of halos that are
separated into nine samples in different mass ranges:
$10.0\le \log M_h <10.5$, $10.5\le \log M_h <11.0$, ...
$14.0\le \log M_h$. The open squares shown in Fig. \ref{fig:ACF1} are
the ACFs we measured from these halos. The error bars are once again
obtained from 200 bootstrap re-samplings of the dark matter halos. We
also show, for comparison, the ACF of dark matter particles as solid
line, and the biased ACF as dotted one. The later is obtained by
multiplying the unbiased ACF by the square of the median bias factor
of those halos in consideration.  Similarly to the CCFs, we find that
the ACFs of halos are quite consistent with the biased ACFs of dark
matter particles on large scales.  The small scale ACFs cut off is
especially apparent for massive halos. These are caused by the
halo-halo exclusion effect as modeled in \citet{Wang2004}.  Overall,
the ACFs in different mass bins behave as expected.

Next, we proceed to measure the ACFs for halos in different cosmic web
environments. Shown in each panel of Fig. \ref{fig:ACF2} are the ACFs
for halos in different cosmic web environments: clusters, filaments,
sheets and voids.  These ACFs show quite different behaviors compared
to the overall sample. This further indicates that a significant
cosmic web environment dependence exists. On small to intermediate
scales, the clustering properties of ACFs are different from CCFs,
especially for the halos in the voids where they show
overall stronger clusterings.  Nevertheless, the clustering behaviors
on these scales are not the main focus of this probe.  The most
exciting feature is that, on scales $\ga 15\mpch$, the halos show the
same cosmic web environment dependence as the CCFs, where strongest
clustering strength is revealed in the ACFs of void halos.

The cosmic web dependence of the clustering strengths found here is
opposite to that obtained by \citet{Fisher2016}, who claimed that halo
bias monotonically increases with cosmic web environments from voids
to clusters.  Although it is unclear to us what is the main cause of
such a discrepancy, one possibility is that they used the velocity
shear tensor to classify halos into different cosmic web types.
Nevertheless, in a very recent study using a Fourier analysis method,
\citet{Paranjape2017} found the same trends of halo biases as ours:
the halos in filaments have significantly larger biases than those in
clusters.  Coming back to the strongest clustering strength of void
halos, one intuitive explanation is that the clustering of void halos
might be composed of the void-void correlation weighted by the void
occupation numbers.

\begin{figure*}
\center
\vspace{0.5cm}
\includegraphics[height=10.0cm,width=10.0cm,angle=0]{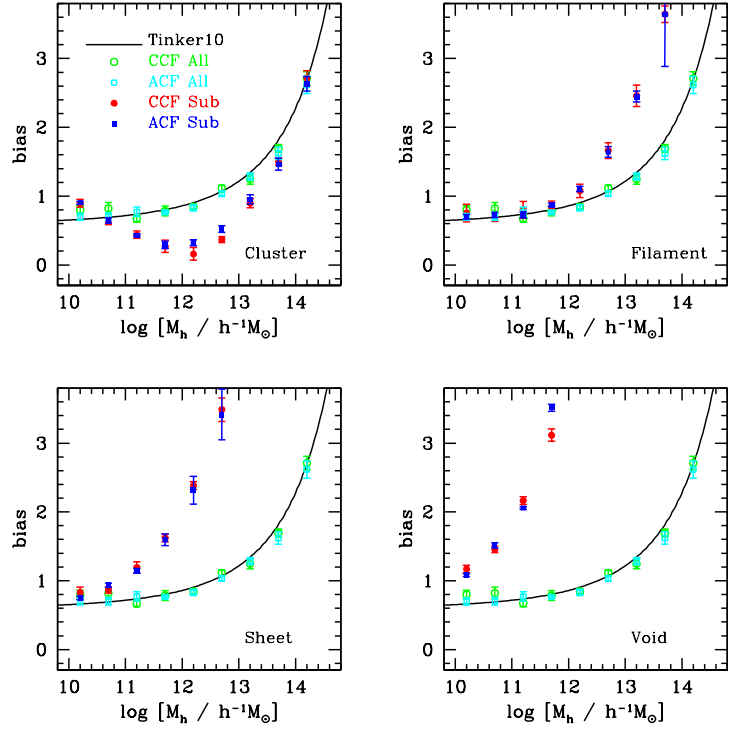}
\caption{Halo biases as a function of halo mass. Each panel refers to
  a different cosmic web environment. Halo biases calculated using the
  CCFs and ACFs of the overall population are shown with open circles
  and squares, respectively. The ones obtained from CCFs and ACFs of
  halo subsamples are displayed as solid circles and squares,
  respectively. For comparison, the model prediction of
  \citet{Tinker2010} is indicated by the solid line. }
\label{fig:bias1}
\end{figure*}
\begin{figure*}
\center
\vspace{0.5cm}
\includegraphics[height=10.0cm,width=10.0cm,angle=0]{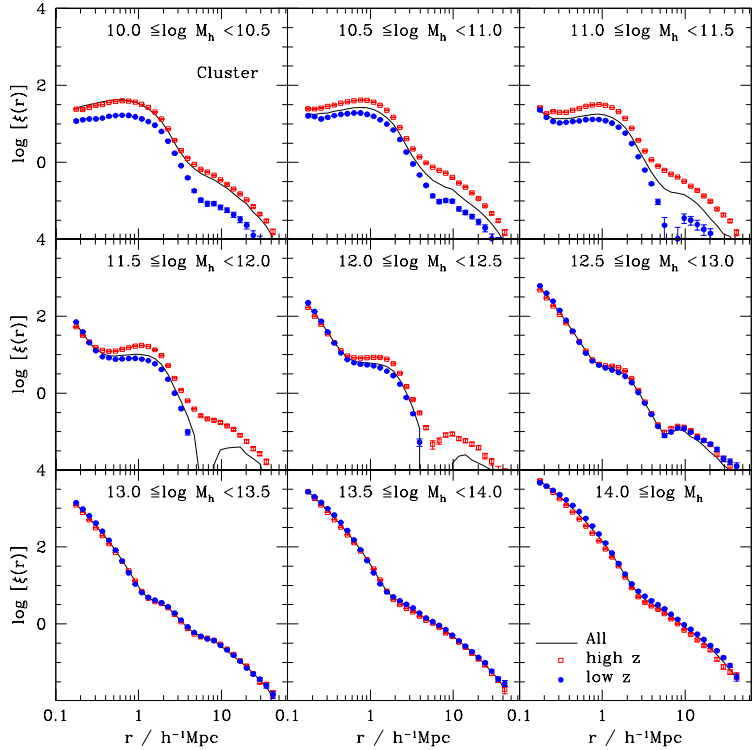}
\caption{Cross correlation functions between halos in the clusters
   and the dark matter particles.  In each panel, the open
  squares and solid dots focuses on the 20\% oldest and 20\% youngest
  halos, respectively. For comparison, we also show using a solid line
  the results for overall population. }
\label{fig:ageCCF1}
\end{figure*}

\subsection{Cosmic web dependent  halo bias}

In order to quantify the overall clustering strengths of halos within
different cosmic web environments, we measure the average biases of
halos. These average biases are measured using the ratios between the
halo CCFs and the dark matter particles' ACFs within radii
$15<r<45\mpch$ as shown in Fig. \ref{fig:CCF1}.  In all panels of
Fig. \ref{fig:bias1} and displayed with open circles are the average
biases thus measured for overall population as a function of halo
mass. The error bars indicate the variances of the CCF ratios (biases)
at different radii. Once again we use the halo ACFs to confirm those
measurements, and display as open squares the average square root
ratios between the ACFs of halos and dark matter particles within
radii $15<r<45\mpch$. This second measurements of the average biases
obtained agree very well with those obtained from the CCFs.  For
completeness, we have added to the figure a solid line representing
theoretical model prediction by \citet{Tinker2010}. The average biases
of halos measured from the ELUCID simulation show an overall very nice
agreement with the model.

In addition to the generic features relevant to the overall population
we have just mentioned, each panel is used to distinguish a specific
cosmic web environment for which the halo bias measurements are
repeated. Environmental bias measurement made using the CCF are shown
as red filled circles.  While the ones made using the ACF are shown as
filled squares.  The average biases of halos in the clusters
(upper-left panel of Fig. \ref{fig:bias1}) agree with the overall
population on both massive and low mass end, but are significantly
suppressed for intermediate mass halos. The average biases of halos
with mass $\sim 10^{12.0}\msunh$ is almost by a factor of 3 lower than
the overall population.  Compared to those in clusters, halos in
filaments (upper-right panel) show significantly enhanced biases,
especially at the massive end with mass $\ga 10^{12.5}\msunh$.  Shown
in the lower-left and lower-right panels of Fig. \ref{fig:bias1} are
the average biases of halos in the sheets and the voids, respectively.
Again we see a clear trend that halos in the sheets and voids are ever
increasingly enhanced.  Overall, the clustering strength of halos
increases significantly from the clusters to the voids, with stronger
impact in more massive halos.  For instance, Milky Way sized halos
(mass $\sim 10^{12.0}\msunh$) in the voids have bias $\ga 3$, which is
even larger than the overall clustering strength of cluster sized
halos.

Finally, for reference, the related biases of halos in different
cosmic web environments are listed in Table \ref{tab:bias}.  Such a
strong cosmic web dependent halo bias may play important roles in
cosmological and galaxy formation probes. For instance,
\citet{Hamaus2014} proposed a method of using galaxy-void cross
correlation to constrain cosmologies, the very different clustering
behaviors of void halos found here must be properly taken into account
in such kind of studies.  On the other hand, as pointed out in ELUCID
III, the fractions of red galaxies as functions of the $r$-band
absolute magnitude depend very strongly on the cosmic web
environments. It would be very interesting to see if galaxies in
different cosmic web environments have different galaxy-halo
connections. For this purpose, the very different halo biases in
different cosmic web environments have to be taken into account in
establishing the galaxy-halo connections using the HOD or CLF models.

\begin{figure*}
\center
\vspace{0.5cm}
\includegraphics[height=10.0cm,width=10.0cm,angle=0]{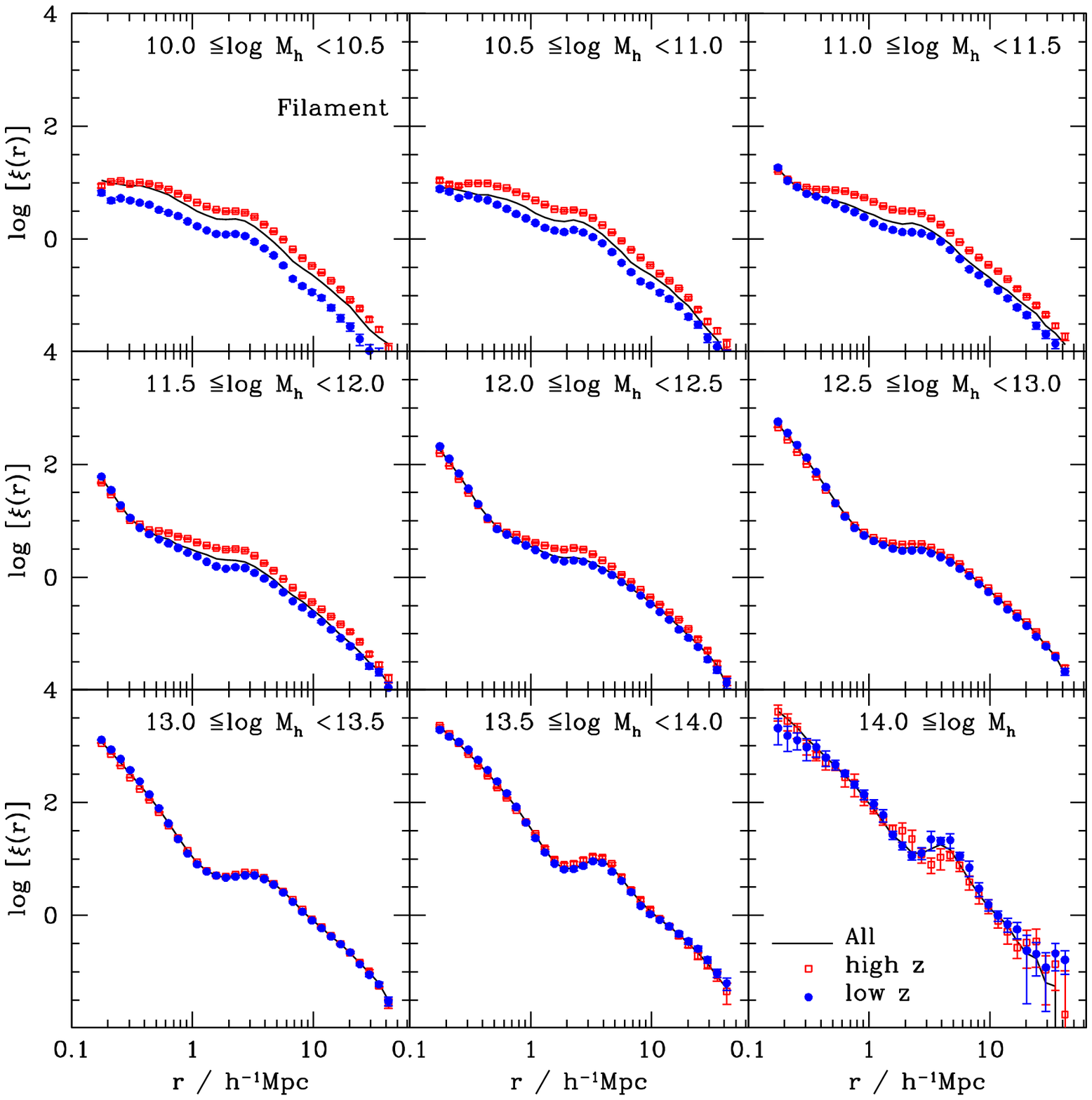}
\caption{Similar to Fig. \ref{fig:ageCCF1}, but for halos in the
  filaments. }
\label{fig:ageCCF2}
\end{figure*}
\begin{figure*}
\center
\vspace{0.5cm}
\includegraphics[height=10.0cm,width=10.0cm,angle=0]{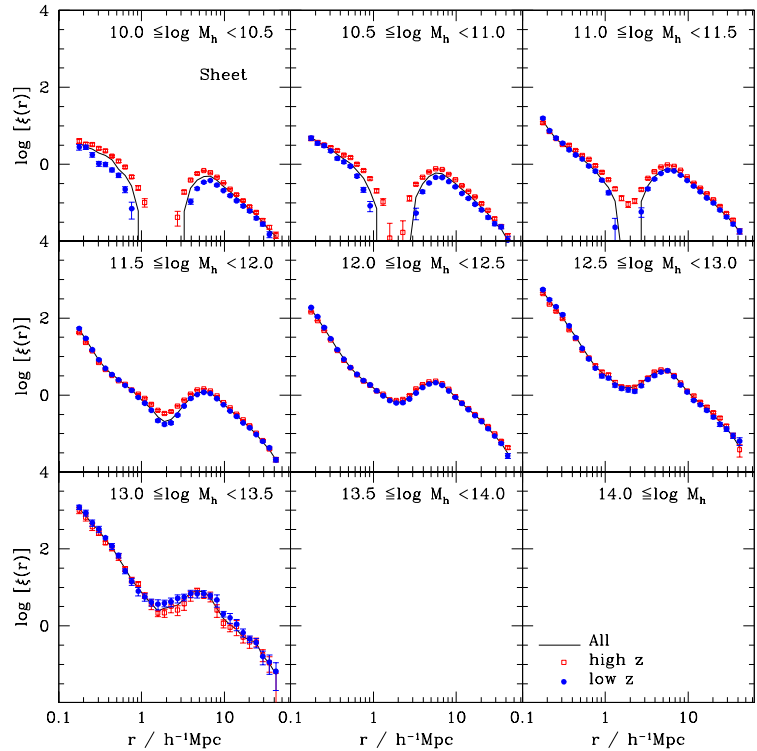}
\caption{Similar to Fig. \ref{fig:ageCCF1}, but  for halos in the
  sheets. }
\label{fig:ageCCF3}
\end{figure*}
\begin{figure*}
\center
\vspace{0.5cm}
\includegraphics[height=7.0cm,width=10.0cm,angle=0]{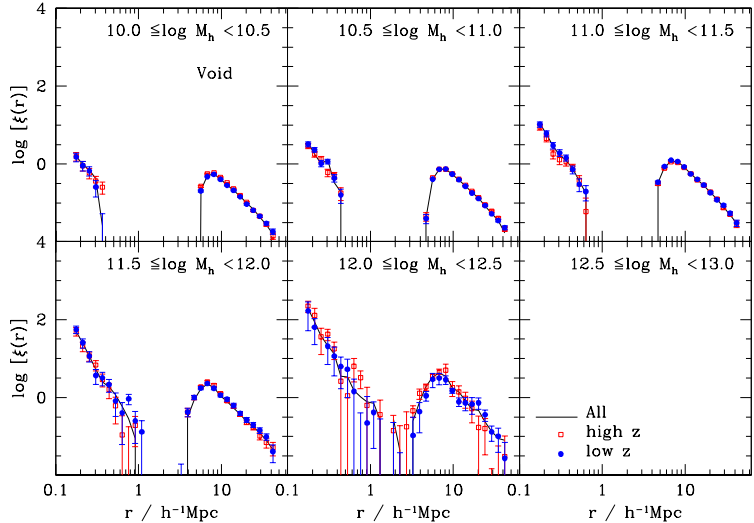}
\caption{Similar to Fig. \ref{fig:ageCCF1}, but  for halos in the
  voids. }
\label{fig:ageCCF4}
\end{figure*}
\begin{figure*}
\center
\vspace{0.5cm}
\includegraphics[height=10.0cm,width=10.0cm,angle=0]{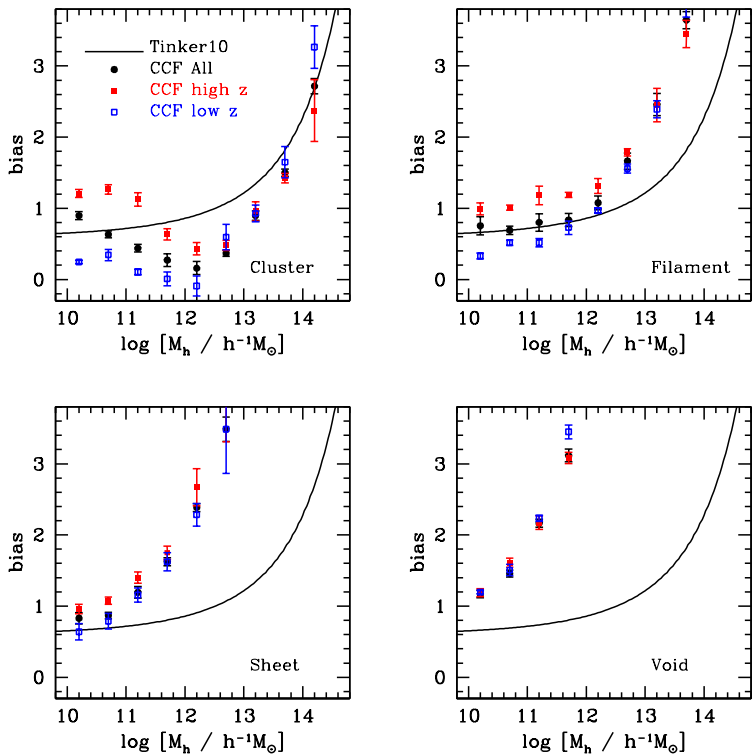}
\caption{Halo biases as a function of halo mass. Each panel represent
  a different cosmic web environment. Halo biases are computed using
  the CCF from all halos (filled black circle), the 20\% oldest
  (filled red squares) and the 20\% youngest (open blue squares)
  halos.  For comparison, we also show the model prediction of
  \citet{Tinker2010} using a solid line.  }
\label{fig:agebias}
\end{figure*}
\begin{figure*}
\center
\vspace{0.5cm}
\includegraphics[height=10.0cm,width=10.0cm,angle=0]{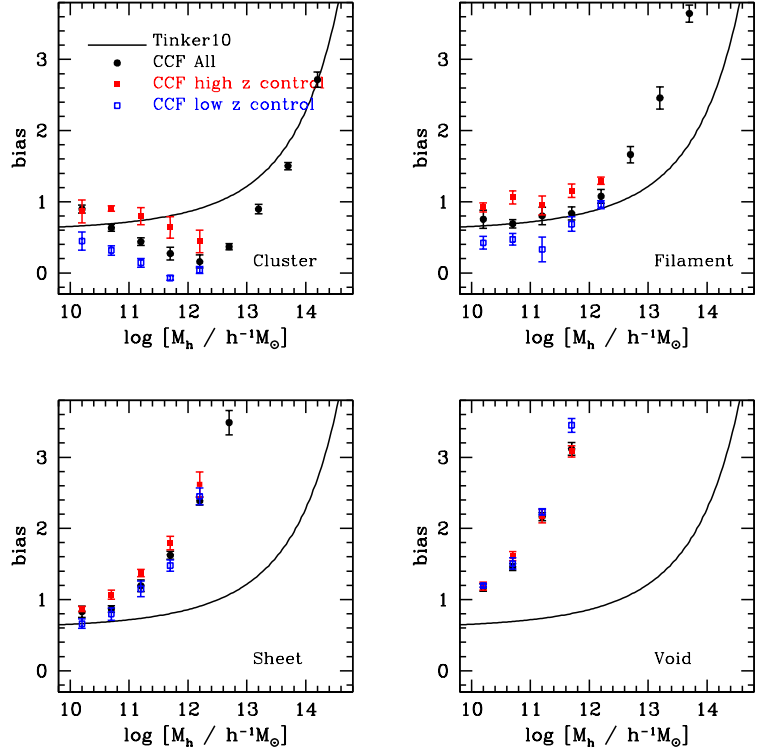}
\caption{Similar to Fig. \ref{fig:agebias}, but here for halos in
    the controlled samples. The results shown in the lower-right panel
    for void halos are the same as those in Fig. \ref{fig:agebias}. Halo
    biases computed in other panels for the 20\% oldest (filled red
    squares) and the 20\% youngest (open blue squares) are obtained
    from the controlled samples. }
\label{fig:agebias2}
\end{figure*}

\section{Assembly bias of halos in different cosmic web
  environments}
\label{sec_age}

In recent years, great effort has been made to quantify the assembly
bias of halos, among which the most notable one is the age dependence.
Having quantified the cosmic web environment dependence of halo
biases, we come to the second purpose of this work: to investigate
whether or not the age dependent biases of halos can be explained
using the cosmic web environment dependent halo biases.  More
precisely we try to confirm the presence or absent of prominent age
dependent biases for halos in different cosmic web environments.  To
this end, we separate halos in the two tails of the formation redshift
distribution, building 2 sets of halo sub-subsamples representing the
20\% oldest and 20\% youngest ones and measure separately their
clustering properties.

We first measure the CCFs of halos in the clusters. The open squares
and solid dots in Fig. \ref{fig:ageCCF1} indicate the resulting CCFs
for the oldest (high z) and youngest (low z) halos. For halos with
mass $\la 10^{12.5}\msunh$, the clustering strengths of the two
sub-subsamples are quite different, the oldest sub-subsample having
much stronger clustering strength than the youngest
sub-subsample. However, for more massive halos, the two CCFs are quite
consistent with one another indicating a lack of assembly bias. As the
assembly bias features found in \citet{Gao2005} are prominent within
mass range $\la 10^{13.5}\msunh$, or
$\la 10^{13.0} \sim 10^{13.5} \msunh$ for different definitions of
formation time \citep[see, e.g.][]{Li2008}, the assembly bias feature
found here is at least limited to a smaller mass range.

Next, we focus on the filaments, where the results are
shown in Fig. \ref{fig:ageCCF2} following the same legend as
Fig. \ref{fig:ageCCF1}. Compared to halos in the clusters,
the assembly bias features differ in two aspects: (1) the overall
strength decreases and (2) the reduction is more significant in more
massive halos.

Finally in Figs. \ref{fig:ageCCF3} and \ref{fig:ageCCF4}, we show the
halo CCFs respectively in the sheets and the voids. Compared to the
clusters and the filaments, assembly bias features are further reduced
in the sheets, especially in terms of amplitude. In the void
environment, halo assembly bias features are no longer apparent, with
the oldest and youngest halos overlapping with the overall void halo
population.

In order to quantify the assembly biases for halos in different cosmic
web environments, we use once again the ratios of the halo CCFs shown
in Figs. \ref{fig:ageCCF1} - \ref{fig:ageCCF4} and the dark matter
ACFs within radii $15<r<45\mpch$.  The average bias measurements are
displayed as a function of mass in Fig. \ref{fig:agebias}, with each
panel corresponding to a specific environment. In each panel, the
biases for overall population, oldest 20\% and youngest 20\% halos are
shown using solid dots, solid squares and open squares, respectively.
The error bars again are obtained using the variances of the CCF
ratios (biases) at different radii.  For the cluster environment in
the upper-left panel, the assembly biases is clearly apparent with the
oldest and youngest halos results being located respectively above and
below the overall cluster population.  The assembly bias of halos in
clusters is found here to be of similar amplitude to the one measured
by \citet{Gao2005}. However it appears in our case in less massive halos
with mass $\la 10^{12.5}\msunh$. As we explore the filament
environment (upper-right panel) we notice a reduction in the amplitude
of the assembly biases. The reduction is even more pronounced in the
sheets (lower-left panel) and the assembly bias entirely absents in
the voids (lower right panel).

Note however, as the halos in different environments have different
age distributions (see Fig. \ref{fig:age}), we carry out an additional
test to check if the reduction of assembly bias, especially that in
voids, is really due to the fixed environment.  Here we take the halos
in the voids as the reference sample and select a set of controlled
halo samples from other environments that match the same mass and
formation time distributions. Following the same procedures we carried
out for the reference sample in voids, we measure the biases of the
oldest 20\% and youngest 20\% halos in the controlled samples in
clusters, filaments, and sheets, respectively. Thus obtained results
are shown in Fig. \ref{fig:agebias2}.  Here only halos with mass
$\le 10^{12.5}\msunh$ are used for our investigation. Compare to the
original halos shown in Fig. \ref{fig:agebias}, the ones in the
controlled samples show very similar, although somewhat less
prominent, assembly bias features.  These features indicate that the
reduction of assembly bias from clusters, filaments to sheets and
voids is real.

Although the results are not displayed in the paper, the measurements
using the ACF alternatives (as in Fig. \ref{fig:bias1}) were performed
and found to give very similar results.

\section{Summary}
\label{sec_conclusion}

In this paper, we have probed the clustering properties of halos in
different cosmic web environments using the ELUCID simulation. Halos
were separated into four cosmic web environments defined through the
Hessian matrix of the smoothed density field.  Out of a total of
$48,129,323$ halos in the ELUCID simulation, $19.8\%$ are located in
cluster, $57.0\%$ in filaments, $22.1\%$ in sheets and $1.1\%$ in
voids.  In order to probe the related mass dependence of the
clustering properties, we also separated the halos into nine halo mass
bins, $10.0\le \log M_h <10.5$, ..., $13.5\le \log M_h <14.0$ and
$14.0\le \log M_h$.  Furthermore, so as to confirm existing age
dependent assembly biases in different cosmic web environments,
additional subsamples were made based on the formation redshift
focussing on the 20\% oldest and 20\% youngest halos.

We have measured the cross correlation functions between halos and
dark matter particles, as well as the related auto correlation
functions of halos in these subsamples. The ratios between the halo
CCFs (or ACFs) and the dark matter ACFs within radii $15<r<45\mpch$
are used to estimate the average biases of halos. We find
significantly different biases for halos of {\it same} masses but in
different cosmic web environments. We summarize our main results as
follows;
\begin{itemize}
\item The average biases of halos in  {\it clusters} agree with the
  overall population on both high and low mass ends, but are
  significantly {\it suppressed} in intermediate mass scales. The
  average biases of halos with mass $\sim 10^{12.0}\msunh$ are almost
  by a factor of 3 lower than the overall population.
\item The halos in  {\it filaments} show significantly enhanced
  biases, especially at the massive end with mass
  $\ga 10^{12.5}\msunh$ .
\item Halos in  {\it sheets} show stronger enhanced biases as a
  function of halo mass.
\item Halos in   {\it voids} show the strongest bias
  enhancements. The Milky Way sized halos with mass
  $\sim 10^{12.0}\msunh$ in void region have bias $\ga 3$, which is
  even larger than the overall clustering strength of cluster sized
  halos.
\item The strength of the assembly biases (age dependences)
    decreases from cluster, filament and sheet environments to finally
    disappears in void environment. 
\item In addition, the age dependent assembly biases of halos in
    different cosmic web environments exist in a smaller mass range
    $\la 10^{12.5}\msunh$, compared to the overall population at
    $\la 10^{13.0}\sim 10^{13.5}\msunh$. 
\end{itemize}

As we have found very prominent -- as strong as the mass dependence --
cosmic web environment dependence of halo biases, a similar study
focussed on galaxy clustering in different cosmic web environments
would be relevant. It would indeed be interesting to obtain the
galaxy-halo connections in different cosmic web environments, which
can be used to better constrain galaxy formation models and better
understand the void-galaxy clustering properties for cosmological
studies, etc.


\section*{Acknowledgments}

We thank the anonymous referee for helpful comments that greatly
improved the presentation of this paper.  This work is supported by
the 973 Program (No. 2015CB857002), the national science foundation of
China (grant Nos. 11233005, 11621303, 11522324, 11421303) and a grant
from Science and Technology Commission of Shanghai Municipality
(Grants No. 16DZ2260200).  We also thank the support of the Key
Laboratory for Particle Physics, Astrophysics and Cosmology, Ministry
of Education, and Shanghai Key Laboratory for Particle Physics and
Cosmology (SKLPPC)

We acknowledge the computing facility (High Performance Computing
Center) at Shanghai Jiao Tong University for providing access to the
PI cluster on which the ELUCID simulation was run. This work is also
supported by the High Performance Computing Resource in the Core
Facility for Advanced Research Computing at Shanghai Astronomical
Observatory.


\label{lastpage}

\end{document}